\documentclass[runningheads]{svmult}

\usepackage{makeidx}   
\usepackage{graphicx}  
\usepackage{subeqnar}  
\usepackage{multicol}  
\usepackage{physprbb}  
\makeindex             

\begin{document}
\title*{Selection of extragalactic Targets for AO and VLTI
Observations}
\toctitle{Selection of extragalactic Targets for AO and VLTI
Observations}
%
%
\titlerunning{Selection of extragalactic Targets for AO and VLTI}
%
\author{Jens Zuther \inst{1,2}
\and Andreas Eckart\inst{1}
\and Wolfgang Voges\inst{2}
\and Thomas Bertram\inst{1}
\and Christian Straubmeier\inst{1}
}
\authorrunning{J. Zuther et al.}
%
%
\institute{1. Physikalisches Institut, Universit\"at zu K\"oln, K\"oln, Germany
\and Max-Planck-Institut f\"ur extraterrestrische Physik, Garching, Germany
}

\maketitle              

\begin{abstract}
In this contribution we would like to emphasize the usefulness of data mining multiwavelength surveys like the Sloan Digital Sky Survey (SDSS) or the ROSAT All Sky Survey (RASS) -- which have become available to the public recently -- in order to find interesting objects suitable for adaptive optics (AO) or interferometric (VLTI) observations in the infrared. We will present a sample of extragalactic X-ray sources having an optical counterpart (based on SDSS data release 1) which are suitable for AO/VLTI observations using a natural guide star in their vicinity.
\end{abstract}

\section{Introduction}
A major cornerstone for the future of ground-based observations is the availability of adaptive optics (AO) systems on large telescopes (for reviews see \cite{beck93} and \cite{quirr01}). With AO one can overcome the limitations imposed by the earth's atmosphere on image quality in terms of resolution and sensitivity. The result is imaging and spectroscopy at or close to the diffraction limit of the telescope\footnote{For example the ESO Very Large Telescope (VLT) with its 8~m primary mirrors provides a diffraction-limited resolution of about 50~mas at 1.65~$\mu$m using the AO system NACO \cite{brand02}}. For AO observations a natural guide star (NGS) is needed as a reference source to assess the degradation of the wavefronts due to the turbulent atmosphere. The availability of a bright enough reference source significantly reduces the sky coverage. However, large scale surveys like the Sloan Digital Sky Survey\footnote{Web site: \textsf{www.sdss.org}} (SDSS, \cite{york00}) and the ROSAT All Sky Survey (RASS, \cite{voges99}) provide means to effectively search for interesting extragalactic sources suitable for AO observations. 

\section{The X-Ray Background}
A challenge for large telescopes like the VLT is to study faint optical counterparts of X-ray background (XRB) objects in order to understand e.g. the connection between nuclear activity (accretion onto a supermassive black hole) and structure of the host galaxy (e.g. morphology, stellar populations, etc.), i.e. to find out about the physical conditions of active galactic nuclei (AGN) and the galaxy hosts they reside in.

The soft XRB in the 0.5-2~keV regime has been mostly resolved into discrete sources, of which the majority turns out to be AGN. Their nature still remains  mysterious. What is found is that the XRB is much harder than the X-ray emission of unobscured AGN in the local Universe. Therefore the existence of a substantial obscured AGN population is required. However, the question remains whether these objects are hardened due to extinction or whether the central engine itself has a hard spectrum. The hard X-ray sources are found preferentially at lower redshifts, in contradiction to predictions of XRB models \cite{gilli03}.
NIR observations of optical counterparts within reach of 10~m class telescopes can contribute to these issues by providing intrinsic host galaxy luminosities as well as information on the host environment (nuclear excitation, star formation as well as dynamically deduced host and possibly black hole masses) (cf. \cite{zuther03a}). 

\begin{figure}[ht!]
\begin{center}
\includegraphics[width=10cm]{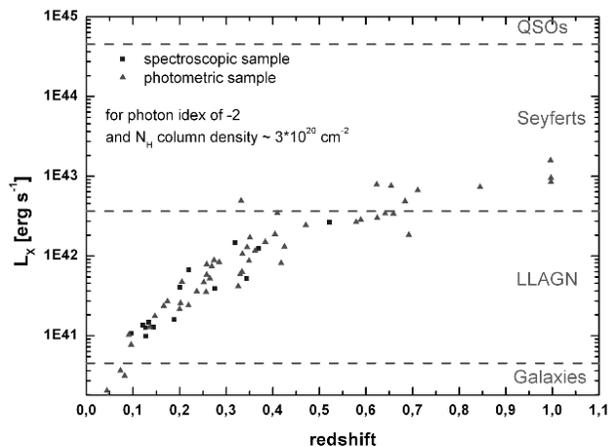}
\end{center}
\caption[]{Hubble diagram ($L_{0.5-2\mathrm{keV}}$/redshift) for the sample. Different source types are indicated on the right axis (Hasinger, private communication). Filled squares correspond to spectroscopic, filled triangles to photometric redshifts \cite{csabai03}.}
\label{fig:1}
\end{figure}
\section{Our X-Ray/optical Sample}
Based on a cross-correlation of optical (SDSS, first data release \cite{dr1}) and X-ray data (RASS) we searched for extragalactic targets having a possible NGS within an angular distance of $40''$ and a $r$-band\footnote{Throughout this article we use the SDSS magnitude system \cite{fuk96} $u, g, r, i$, and $z$.} brightness $<15$. Together with the Petrosian magnitude, $r_\mathrm{Petrosian}$ \cite{edr}, of the galaxy being $r_\mathrm{Petrosian}<20$ we make sure that the sample allows for reasonable AO performance. With the $K$-band fringe tracker and PRIMA, the presented extragalactic objects will also be well in reach of future VLTI observations (e.g. \cite{quirr03}).
This straightforward search resulted in 317 candidate pairs. The objects classified as galaxies suffer a certain contamination by saturated stars, which we addressed by visual inspection. 

The next step, as discussed in \cite{zuther03a}, was to make sure we only take those galaxies which are the most probable counterparts of the X-ray sources, i.e. (1) a X-ray/optical angular separation of less than $40''$, (2) a hardness ratio larger than $-1$, and (3) no X-ray extent of the X-ray source with respect to the ROSAT point-spread function. Application of these criteria gives a final set of 78 galaxy/NGS pairs.
The galaxies of this sample cover the redshift range $0.001<z<1$ (Fig. \ref{fig:1}).
Using the eXsas tool, we calculated the X-ray luminosities assuming a power-law index of $-2$ and an average galactic $N_{\mathrm{H}_2}$ column density of $3\times 10^{20}$~cm$^{-2}$. The X-ray luminosities are in the range $7\times  10^{33}<L_{0.5-2\mathrm{keV}}<2\times 10^{43}$~erg~s$^{-1}$ (Fig. \ref{fig:1}).
Studying the optical colors ($u-g$ and $g-r$), our manual classification is consistent with classification schemes like the $u-r=2.2$ color separator between late- and early-type galaxies \cite{strateva01}, or typical QSO colors (where $u-g<0.6$) \cite{rich01}, and the classification with respect to the X-ray luminosity (Fig. \ref{fig:1}).

Our sample therefore represents a significant first step to a statistically relevant -- i.e. large number of interesting sources -- sample of galaxies suitable for AO and interferometric observations.
\section{Summary}
The combination of the SDSS and ROSAT surveys demonstrates to be a rich source for the search for interesting extragalactic targets. This is especially useful for future high resolution and sensitive imaging and spectroscopy in the infrared on large telescopes like the VLT(I) and the LBT. 

%


\begin{thebibliography}{8.}
\addcontentsline{toc}{section}{References}
\bibitem{beck93} J. M. Beckers: ARA\&A, \textbf{31}, 13 (1993)
\bibitem{quirr01} A. Quirrenbach: ARA\&A, \textbf{39}, 353 (2001)
\bibitem{brand02}W. Brandner et al.: The Messenger, \textbf{107}, 1 (2002)
\bibitem{york00}D. G. York et al.: AJ, \textbf{120}, 1579 (2000)
\bibitem{voges99}W. Voges et al.: A\&A, \textbf{349}, 389 (1999)
\bibitem{dr1} K. Abazajian et al.: AJ, \textbf{126}, 2081 (2003)
\bibitem{fuk96}M. Fukugita et al.: AJ, \textbf{111}, 1748+ (1996)
\bibitem{edr}C. Stoughton et al.: AJ, \textbf{123}, 485 (2002)
\bibitem{gilli03} R. Gilli: astro-ph/0303115 (2003)
\bibitem{zuther03a} J. Zuther: 'A Sample of X-Ray active extragalactic Sources suitable for NIR Adaptive Optics Observations'. In: \emph{AGN Physics with the Sloan Digital Sky Survey}. ed. by G. T. Richards and P. B. Hall (San Francisco: ASP, 2004); astro-ph/0310371
\bibitem{quirr03}A. Quirrenbach: Ap\&SS, \textbf{286}, 277 (2003)
\bibitem{strateva01} I. Strateva et al.: AJ, \textbf{122}, 1861 (2001) 
\bibitem{rich01} G.~T. Richards et al.: AJ, \textbf{121}, 2308 (2001) 
\bibitem{csabai03}I. Csabai et al.: AJ, \textbf{125}, 580 (2003)
\end{thebibliography}
\end{document}